\documentclass[conference]{IEEEtran}
\usepackage{amsmath}
\usepackage{multicol}
\usepackage{amssymb}
\usepackage{graphicx}
\usepackage{algcompatible}
\usepackage{afterpage,lipsum}
\usepackage{supertabular}
\usepackage{longtable}
\usepackage{multirow}
\usepackage{pbox}
\usepackage{url}
\usepackage{lscape}
\usepackage{verbatim} 
\usepackage{array}
\usepackage{adjustbox,lipsum}
\usepackage[ruled,vlined,linesnumbered]{algorithm2e}
\usepackage{caption}
\usepackage{subcaption}
\usepackage{footnote}

\begin{document}

\title{Anvaya: An Algorithm and Case-Study on Improving the Goodness of Software Process Models generated by Mining Event-Log Data in Issue Tracking Systems 
}
\author{\IEEEauthorblockN{ Prerna Juneja\IEEEauthorrefmark{1},
Divya Kundra\IEEEauthorrefmark{1} and
Ashish Sureka\IEEEauthorrefmark{2}}
\IEEEauthorblockA{\IEEEauthorrefmark{1}Indraprastha Institute of Information Technology, Delhi (IIIT-D), India\\
prerna1399@iiitd.ac.in, divya1395@iiitd.ac.in}
\IEEEauthorblockA{\IEEEauthorrefmark{2}Software Analytics Research Lab (SARL), India\\
ashish@iiitd.ac.in}}

\maketitle

\begin{abstract}
Issue Tracking Systems (ITS) such as Bugzilla can be viewed as Process Aware Information Systems (PAIS) generating event-logs during the life-cycle of a bug report. Process Mining consists of mining event logs generated from PAIS for process model discovery, conformance and enhancement. We apply process map discovery techniques to mine event trace data generated from ITS of open source Firefox browser project to generate and study process models. Bug life-cycle consists of diversity and variance. Therefore, the process models generated from the event-logs are spaghetti-like with large number of edges, inter-connections and nodes. Such models are complex to analyse and difficult to comprehend by a process analyst. We improve the Goodness (fitness and structural complexity) of the process models by splitting the event-log into homogeneous subsets by clustering structurally similar traces. We adapt the K-Medoid clustering algorithm with two different distance metrics: Longest Common Subsequence (LCS) and Dynamic Time Warping (DTW). We evaluate the goodness of the process models generated from the clusters using complexity and fitness metrics. We study back-forth \& self-loops, bug reopening, and bottleneck in the clusters obtained and show that clustering enables better analysis. We also propose an algorithm to automate the clustering process -the algorithm takes as input the event log and returns the best cluster set.
\end{abstract}

\IEEEpeerreviewmaketitle
\section{Research Motivation and Aim}
Software Process Intelligence (SPI) is an emerging and evolving discipline involving mining and analysis of software processes. This is modeled on the lines of application of Business Intelligence techniques to business processes (Business Process Intelligence (BPI)), but with the focus on software processes and its applicability to Software Engineering (SE) and Information Technology (IT) systems. SPI has diverse applications and is an area that has recently attracted several researcher's attention due to availability of vast data generated during software development. Some of the business applications of process mining on software repositories or SPI are: uncovering runtime process models, discovering process inefficiencies and inconsistencies, observing project key indicators and computing correlation between product and process metrics, extracting general visual process patterns for effort estimation and analyzing problem resolution activities \cite{poncin2011process} \cite{rubin2007process}.

%\afterpage{%
\begin{figure*}
\centering
\noindent\includegraphics[width=16cm,height=8cm]{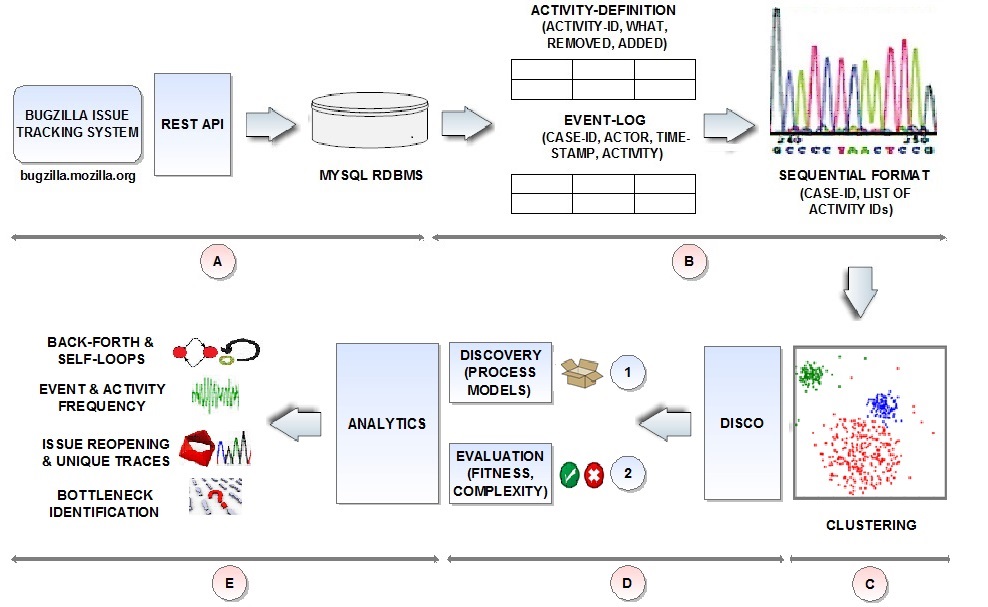}
\caption{Architecture Diagram and Data Processing Pipeline for Anvaya Framework (Clustering-Based Approach for Improving the Goodness of Software Process Models Derived from Event-Logs).}
\label{fig:framework}
\end{figure*}
%} % end of "afterpage" group

Several SE processes such as issue or defect resolution are flexible and consists of several process variants (that are adhoc and unstructured) and a wide spectrum of behavior. This results in a spaghetti process model consisting of a large number of activity or task nodes as well as a large number of relations (or directed edges) between these nodes. A spaghetti process model is structurally complex and hard to comprehend for a process analyst. Trace clustering is a technique which has been applied on business process logs to split a given event-log into homogenous subsets from which process models are uncovered. Trace clustering has shown to improve the comprehensibility of process models in environments which allow process flexibility and large number of variants. The research motivation of the study presented in this paper is to investigate the application of trace clustering in the domain of SPI and process mining software repositories. The specific research aim of the work presented in this paper are the following:
\begin{enumerate}
\item To study the problem of spaghetti process models in the domain of software defect and issue resolution by conducting a case-study on open-source Firefox browser project.
\item To propose a trace clustering technique based on grouping sequential data and apply it on issue tracking system dataset of a large, complex and log-lived open-source project. To investigate the effectiveness of the proposed trace clustering technique in reducing the structural complexity and enhancing the process model comprehensibility for a process analyst.
\item To study self-loops, back-and-forth, issue reopen and bottlenecks on the discovered process models from the homogeneous subset output of trace clustering and illustrate its benefits in the domain of SPI using a real-life case-study.
\end{enumerate}

\begin{table*}[t]
\begin{flushleft}
\large
\begin{flushleft}
\caption{Count and Description of 11 out of 81 Unique Activities in the Experimental Dataset.}
\label{table:activities}
\end{flushleft}
\begin{adjustbox}{width=\textwidth,totalheight=\textheight,keepaspectratio}
%\begin{tabular}{|p{0.3 cm}|p{5 cm}|p{1 cm}|p{10 cm}|}

\begin{tabular}{|m{4 cm}| 
 >{\centering\arraybackslash}p{1 cm}| >{\centering\arraybackslash}p{1 cm}|p{10 cm}|}

\hline
\multicolumn{1}{|c|}{\textbf{{\footnotesize  Activity}}} & \multicolumn{1}{|c|}{\textbf{{\footnotesize Acronym }}} &
\multicolumn{1}{|c|}{\textbf{{\footnotesize  Count}}} &
\multicolumn{1}{|c|}{\textbf{{\footnotesize  Description}}} \\
\hline

% {\footnotesize  Alias } & {\footnotesize  ALI} & {\footnotesize  $35$ } &
%{\footnotesize  Short name assigned to bug for referring it at other places in Bugzilla.} \\\hline
 {\footnotesize  Assigned To } & {\footnotesize  ASS } &{\footnotesize  $4274$ } & {\footnotesize  Bug is assigned to the resolver by the triager. } \\
\hline
%{\footnotesize  $3$ } & {\footnotesize  Attachments} &{\footnotesize  $7076$ } &{\footnotesize  They carry some information about the bug that is added to it. }\\\hline
%{\footnotesize  $3$ } & {\footnotesize  Blocks (BLO)) } & {\footnotesize  $4730$} &{\footnotesize  Bug to be resolved before the bugs listed in "added" field are resolved.}\\\hline
 {\footnotesize  Carbon Copy } & {\footnotesize  CCC } &{\footnotesize  $48387$ } &{\footnotesize  Adding/Removing people in addition to Reporter, Resolver and QA Contact to the CC list of the bug in order to notify them about the bug's progress.}\\
\hline
{\footnotesize  Custom Field Blocking} & {\footnotesize  CFB } & {\footnotesize   $573$ } &{\footnotesize  Nominating the bug to stop a release by setting the appropriate blocking flag\textsuperscript{\ref{footnote-label}}. }\\
\hline
 
{\footnotesize  Is Confirmed} & {\footnotesize  ISC } & {\footnotesize  $1106$ } &{\footnotesize  Confirming the bug to be true i.e. issue raised is actually a bug.}\\
\hline

{\footnotesize  QA Contact Assigned} & {\footnotesize  QAC} & {\footnotesize  $1271$ } &{\footnotesize  Contacting Quality Assurance agent for either confirming the bug or verifying the fix.  }\\

\hline

 {\footnotesize Status New Resolved } & {\footnotesize SNR } &{\footnotesize  $4492$} &{\footnotesize  The bug status changes from New where it was processed to Resolved where resolution has been performed and is awaiting verification by Quality Assurance. } \\
\hline

{\footnotesize  Status Resolved Reopened } & {\footnotesize  SRR } &{\footnotesize  
$702$} &{\footnotesize The bug status changes from Resolved where its resolution was set to Reopened where the bug is reopened as the resolution is found to be incorrect. }
\\
\hline

 {\footnotesize  Status Resolved Verified  } & {\footnotesize  SRV } & {\footnotesize  $731$} &{\footnotesize The bug status changes from Resolved where its resolution was set to Verified where Quality Assurance has looked at the bug and its resolution and agrees that the appropriate resolution has been performed. }\\
\hline
 {\footnotesize  Status Unconfirmed Resolved } & {\footnotesize  SUR } &{\footnotesize  $5334$} &{\footnotesize The bug status changes from Unconfirmed where it was validated whether the bug is true to Resolved where resolution has been set.}\\
\hline
{\footnotesize  Summary Modified} &{\footnotesize  SUM } &  {\footnotesize  $2362$} &{\footnotesize The short sentences describing what the bug is about are added/removed. }\\
\hline

{\footnotesize  Target Milestone Defined} & {\footnotesize  TAR } & {\footnotesize  $3787$ } &{\footnotesize  Setting the milestone field while the bug is open to indicate the release for which the fix is planned. }\\
\hline
\end{tabular}
\end{adjustbox}
\end{flushleft}
\end{table*}

\section{Reseach Framework and Solution Approach}
Figure \ref {fig:framework} shows the architecture diagram and the $4$ step data processing pipeline for the Anvaya Framework. The first step consists of extracting Issue Tracking System (ITS) data for the Firefox project using the Bugzilla REST API (an HTTP version of its XMLRPC and JSONRPC APIs)\footnote{ https://wiki.mozilla.org/Bugzilla:REST\_API} and saving it in a MySQL Database. We extract the complete history (life-cycle) of all closed bugs. The history consists of five fields: Who, When, What, Removed and Added. An event in an event-log for a process model discovery algorithm requires a minimum of four fields: Case ID (or the Trace ID for the process instance), Actor, Timestamp and Activity. We map the ITS Issue ID as the Case ID, Who as Actor, When as Timestamp and a combination of What, Removed and Added as Activity.

We convert the history into an Event-Log table consisting of three columns [Case Id, Timestamp and Activity] where Activity column consists of the Activity-ID corresponding to What, Added and Removed in the Activity-Definition Table \ref{table:activities}. We extract, label and output all the unique activities from the Bugzilla history into the Activity-Definition Table \ref{table:activities}. For labelling, we use a three letter code which reflects and indicates the activities performed. We identify $81$ unique activities in our dataset. Due to limited space, we present the count and description of only $11$ unique Activity-IDs in Table \ref {table:activities}. We structure the Event-Log data in increasing order of Case IDs and activities within a case instance in increasing order of timestamp. We transform the data into a sequential format since we are applying sequential data clustering. We adapt the $K$-medoid algorithm  to cluster the sequential data using two different distance metrics: Longest Common Subsequence (LCS) and  Dynamic Time Warping (DTW). Output of this step is a set of $k$ clusters. The clustering algorithms are explained in Section \ref{sec:clustering}. We generate a single process model from the entire event-log data as well as for each cluster obtained in the previous step using a process mining tool Disco\footnote{Disco is a process mining toolkit for which we obtained the academic license.} that uses the fuzzy miner algorithm \cite{Gunther:2007:FMA:1793114.1793145}. We choose Disco because of its ability to manage large event logs and produce complex models. We evaluate the goodness of these process models using cyclomatic complexity and fitness metrics. The last step of Anvaya framework is the Analytics Step where we study and mine useful information from the process models generated from the clusters and show benefits of trace clustering in analysis of back-forth \& self loops, bug reopening, and bottlenecks.

\section{Experimental Dataset}
\begin{table}[!htp]
\begin{center}
\footnotesize
\begin{center}
\caption{Experimental Dataset Details (Mozilla Firefox Project) }
\label{table:dataset_details}
\end{center}
\begin{tabular}{|l|l|}
\hline
\multicolumn{1}{|c|}{\textbf{{\footnotesize Attribute}}} & \multicolumn{1}{|c|}{\textbf{{\footnotesize Value}}}  \\
\hline
{\footnotesize Project} & {\footnotesize Firefox}\\
\hline
{\footnotesize First Issue Report Date} & $1$ {\footnotesize January $2013$}\\
\hline
{\footnotesize Last Issue Report Date} & $31$ {\footnotesize December $2013$}\\
\hline
{\footnotesize Data Extraction Date} & $16$ {\footnotesize October $2014$}\\
\hline
%{\footnotesize Issues not Authorized for Access} & {\footnotesize %$156$} \\
%\hline

%{\footnotesize Number of Unique Components of Firefox Used} & %{\footnotesize $60$ (All Included)} \\
%\hline
%{\footnotesize Total Issues} & {\footnotesize $ $} \\
%\hline
{\footnotesize Number of Open Issues} & {\footnotesize $3399 $} \\
\hline
{\footnotesize Number of Closed Issues Used} & {\footnotesize $11804$} \\
\hline
{\footnotesize Number of Activities in Closed Issues} & {\footnotesize $81$}\\
\hline
{\footnotesize Number of Events Reported for Closed Bugs} & {\footnotesize $178331$} \\
\hline
\end{tabular}
\end{center}
\end{table}

We extract close bug report data for Firefox Browser because closed bugs have completed their lifecycle. We do not analyse open bug report data because such bugs are still in the pipeline, work is being done on them, and we don't know what shape they are going to take. The lifecycle of a bug consists of several stages. The initial status of the bug is either New or Unconfirmed. From any of these two states it can either go to Assigned state where it is assigned to a resolver by the triager or can be directly Resolved.  A bug can have seven resolutions: Wontfix, Worksforme, Invalid, Fixed, Remind, Duplicate and Later\footnote{\label{footnote-label}https://bugzilla.mozilla.org/page.cgi?id=fields.html}. Here onwards, the bug is often Verified and Closed or can be Reopened. A bug is said to be closed if its status has been set to either Verified or Resolved. Table \ref{table:dataset_details} shows the experimental dataset details for the Mozilla Firefox project. We conduct experiments on publicly available dataset so that our approach or results can be replicated and used for benchmarking and comparison. We share our dataset and associated files by creating a public repository on GitHub\footnote{https://github.com/ashishsureka/anvaya}
%\clearpage
\section{Clustering}
\label{sec:clustering}
We adapt the $K$-medoid clustering algorithm \cite{kaufman1987clustering} \cite{kmedbook90} to cluster the sequential traces using two different distance metrics. The first distance metric that can be used to compute the similarity between two traces is the Longest Common Subsequence metric (LCS Similarity) \cite{Wagner:1974:SCP:321796.321811} \cite{Bergroth:2000:SLC:829519.830817} \cite{Hirschberg:1977:ALC:322033.322044}. Since each trace can be viewed as a sequence of characters, we use the LCS algorithm to determine the length of the longest common sequence of characters which need not be consecutive but follow a left to right ordering. Longer the length of LCS, more similar will be the traces. The second distance metric we use is Dynamic Time Warping (DTW Similarity) \cite{1275947} \cite{conf/kdd/BerndtC94} which is used to find similarity between sequences that are structurally similar but can be on a different timescale. Let two sequences be S1 and S2. Warping path consists of index pairs (i,j) if DTW associates S1[i] with S2[j]. This path is subjected to certain restrictions namely, monotonicity, continuity
and boundary condition \cite{books/daglib/0018897}. Out of the many warping paths, an optimal warping path is the one that minimizes the total cost \cite{books/daglib/0018897}. Warping distance is the summation of element wise distance between S1[i] and S2[j] over all pairs of (i,j) present in the optimal warping path\footnote{http://cs.bc.edu/\~alvarez/Algorithms/Notes/dtw.html\label{dtwlink1}}. We assign a cost (distance) 0 if  S1[i]=S2[j], otherwise 1 is assigned. Because of such cost assignment, lower the warping distance, more similar are the traces. So, in $k$-medoid algorithm a non medoid trace is associated to a medoid with highest LCS similarity or lowest DTW similarity. Algorithm \ref{algo:kmed} describes the steps to compute $k$ clusters using our proposed technique.

\begin{algorithm}
%\dontprintsemicolon
\KwData{Event log in sequential data format}
\KwResult{$k$ clusters}
\textnormal{input the value of number of clusters to be formed $k$.}\\
\textnormal{read the input event log}\\
\textnormal{randomly select $k$ traces as initial medoids.}\\
%\textnormal{match each two sequences $s1$ and $s2$ to closest medoid using either LCS Similarity or DTW Similarity.} \\ 
\ForEach{\textnormal{ non medoid trace $t_i$}}{
\ForEach{\textnormal{medoid trace $m_i$}}{
\textnormal{calculate similarity score of $t_i$ and $m_i$  using LCS Similarity $lcs_i$ or DTW Similarity $dtw_i$ }

}
\textnormal{assign $t_i$ to $m_i$ with highest $lcs_i$ or lowest $dtw_i$. }

}

\ForEach {\textnormal{medoid trace} $m$ }{
\ForEach{\textnormal{non medoid trace $o$} }{
\textnormal{swap} $m$ \textnormal{and} $o$ \\
\textnormal{compute the total similarity score (cost) of the configuration using either $lcs_i$ or $dtw_i$}} }
\textnormal{select the configuration with the highest cost while using LCS Similarity and lowest cost while using DTW Similarity}.\\
\textnormal{Steps 4 to 12 are repeated till there is no change in the medoids}\\
\caption{$k$ Medoid Clustering }
\label{algo:kmed}
\end{algorithm}
%\clearpage

\section{Process Discovery \& Evaluation}
\begin{figure*}[t]
 \centering
 \begin{subfigure}[b]{0.3\textwidth}
 \includegraphics[width=5cm,height=7cm]{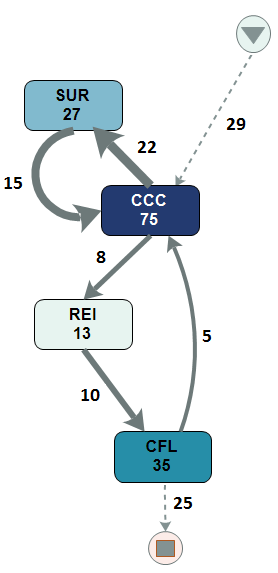}
 \caption{A Process Model example }
 \label{fig:example}
 \end{subfigure}
 \hspace{1cm}
  \centering
 \begin{subfigure}[b]{0.5\linewidth}
 \includegraphics[width=10cm,height=7cm]{main_model.pdf}
 \caption{Main Model}   
 \label{fig:main_model}
 \end{subfigure} 

  \caption{(a) A Process Model generated from Disco with labels as Absolute Frequency and the arrow thickness \& node color proportionate to this frequency. (b) Complex Process Model generated from the the entire event log consisting of 1615 traces.}
  \label{fig:clusters}
\end{figure*}

We discover process models from the entire event log as well as the event log of the clusters using Disco. A node in the process model obtained from Disco represents an Activity while an edge represents transition from one activity to another. The process model has a starting node (represented by a triangle symbol), end node (represented by a stop symbol) and activity nodes containing the name and absolute frequency of the activity. Dashed arrows point to activities that occur at the very beginning or very end of the processes. Transitions between activities are represented by solid directed arrows with the absolute frequency value written over them. The color of nodes and thickness of edges is proportional to their frequency. Darker shade and larger thickness signifies a higher frequency count. Figure \ref{fig:example} shows a process model generated from Disco.

We evaluate the goodness of process models using two metrics defined in the field of process mining, namely complexity and fitness. Process models discovered from clusters should exhibit low degree of structural complexity and high-degree of fitness.

\subsection{Complexity}
Complexity has unwanted effects on understandability, comprehensibility and correctness of process models \cite{Cardoso:2006:DCP:2135571.2135586}. Out of the many complexity metrics proposed in literature, we use McCabe's cyclomatic number which represents the total number of independent paths possible in the process model \cite{McCabe:1976:CM:800253.807712}. The pseudocode to determine the cyclomatic number of process models obtained from Disco is given in Algorithm \ref{complexity}. The Xml format input of the process model is needed as it carries all the relevant information namely, the number of edges and nodes which is required for calculating the complexity. The higher the complexity value returned by this algorithm, higher will be number of independent paths and thus more complex will be the model.

\begin{algorithm}[!htp]
%\dontprintsemicolon
\KwData{Xml format input of the process model}
\KwResult{Complexity of the process model}
\textnormal{read number of edges $e$}\\
\textnormal{read number of nodes $n$}\\
complexity=$e$-$n$+2 \\
\caption{Complexity}
\label{complexity}
\end{algorithm}

\begin{algorithm}
%\dontprintsemicolon
\KwData{Xml format input of the process model and Event log in sequential format.}
\KwResult{Fitness of the process model.}
\textnormal{read Xml format input file.}\\
\ForEach{ \textnormal{transition between a source $n_i$ and target node $n_j$}  } 
{
 adjacency matrix $a_{n_i,n_j}$ =1 \\ }

\textnormal{read the input event log \\}
\ForEach{\textnormal{bug id $b_{i}$ }}
{ add each activity to trace $t_i$ \\ 
\If { $t_i$ is unique } 
{  add  it to $uiquetrace[]$ \\
Count its frequency $F_i$ in the event log\\ }
}
\ForEach{ \textnormal{entry $t_i$ in $uniquetraces[]$ }}{
$Valid_i$=1 \\
$j$=1 \\
\While{\textnormal{ j$<$length of $t_i$ } }{
\If {  $a_{t_i[j],t_i[j+1]}$  $\ne$ \textnormal{$1$} }
{
$Valid_i$=0 \\
break \\
}
\Else{
$j++$ \\
}

}

}
\ForEach{\textnormal{entry $t_i$ in $uniquetraces[]$}}
{
FreqValidProduct=FreqValidProduct+$F_i$*$Valid_i$\\
FreqSum=FreqSum+$F_i$ \\
}
\textnormal{Fitness=FreqValidProduct/FreqSum}
\caption{Fitness}
\label{fitness}
\end{algorithm}

\subsection{Fitness}
One of the major applications of Process Mining is to determine the gaps between the real world as recorded in the event log and the process model\footnote{http://www.processmining.org/online/conformance\_checker}. The fitness metric is used to determine the extent to which an event log conforms to the process model generated from that log and vice versa \cite{annebpmworkshop}. It can be measured by determining the fraction of traces present in the event log that can be completely replayed by the process model from start to end. The pseudocode to determine the fitness of the process model is given in Algorithm \ref{fitness} \cite{Gupta:2014:NMB:2590748.2590749}. The fitness of a process model can take any value on a scale of $0$ to $1$. Fitness value $1$ (maximum) indicates that the process model is perfectly aligned with the event log while value $0$ (minimum) indicates that the model completely deviates from reality since none of the traces present in the event log are shown in the process model.

\section{Experimental Results}
To validate the clustering, we apply $k$-medoid algorithm using LCS and DTW similarity metrics on $1615$ process-instances and obtain $6$ clusters. Figure \ref{fig:main_model} shows the complex, spaghetti-like, hard to comprehend process model generated from the entire event log (referred as the main model throughout the paper) obtained from Disco at $100\%$ activity and $12.2\%$ path resolution. The complexity and fitness of main model and all the clusters is shown in Table \ref{res}. We see that on an average the complexity in a cluster has been reduced by $40.03 \%$ and $40.96 \%$ while using LCS and DTW metrics respectively clearly showing that clusters are much easier to comprehend and analyse. We notice that process models of $66.67\%$ clusters in case of LCS and $83.34\%$ clusters in case of DTW have a better one to one mapping with the event log and thus show a better fitness value. Throughout our work in further sections, we use LCS distance metric for analysis. 

\begin{table}[!htp]
\begin{center}
\normalsize
\begin{center}
\caption{Cyclomatic Complexity along with Percentage Decrease in Complexity of Clusters (DCC) and Fitness  Metric of the Spaghetti Model Generated from the entire Event Log as well as the Six Clusters Generated by $K$-medoid Algorithm using LCS and DTW as the Distance Metrics}
\label{res}
\end{center}

\begin{tabular}{|c|c|c|c|c|}
\hline
\multicolumn{1}{ |c| }{} &
\multicolumn{2}{ |c| }{LCS} &
\multicolumn{2}{ |c| }{DTW} \\
\hline 

\multicolumn{1}{|p{1cm}|}{{\vskip 0.025pt}\centering\arraybackslash\vspace{-1mm}\textbf{ }}
& \multicolumn{1}{|m{1.5cm}|}{\centering \textbf{\footnotesize Cyclomatic Complexity (DCC)} } 
& \multicolumn{1}{|m{0.75cm}|}{\centering \textbf{\footnotesize Fitness } 
}
& \multicolumn{1}{|m{1.5cm}|}{\centering \textbf{\footnotesize Cyclomatic Complexity (DCC)} } 
& \multicolumn{1}{|m{0.75cm}|}{\centering \textbf{\footnotesize Fitness } 
}
\\\hline
{\footnotesize Main Model} & {\footnotesize 143 (-)}& {\footnotesize  0.017} & {\footnotesize 143 (-)}& {\footnotesize  0.017} 
%& {\footnotesize } 
\\
\hline
{\footnotesize Cluster 1} & {\footnotesize 75 (47.5 \%)}& {\footnotesize  0.178} & {\footnotesize  89 ( 37.7 \%)}& {\footnotesize 0.004 }
%& {\footnotesize } 
\\
\hline
{\footnotesize Cluster 2} & {\footnotesize 82 (42.6 \%)}& {\footnotesize 0.085} & {\footnotesize  93 ( 34.9 \%)}& {\footnotesize 0.059} 
%& {\footnotesize } 
\\
\hline
{\footnotesize Cluster 3} & {\footnotesize 106 (25.8 \%)}& {\footnotesize 0.004}  & {\footnotesize  63 ( 55.9 \%)}& {\footnotesize 0.328} 
%& {\footnotesize } 
\\
\hline
{\footnotesize Cluster 4} & {\footnotesize 96 (32.8 \%)}& {\footnotesize 0.070}  & {\footnotesize 97 (32.1 \%)}& {\footnotesize 0.063} 
%& {\footnotesize } 
\\
\hline
{\footnotesize Cluster 5} & {\footnotesize 83 (41.9 \%)}& {\footnotesize 0.015} & {\footnotesize 78 (45.4 \%)}& {\footnotesize 0.052} 
%& {\footnotesize } 
\\
\hline
{\footnotesize Cluster 6} & {\footnotesize 72 (49.6 \%)}& {\footnotesize 0.208}& {\footnotesize 86 (39.8 \%)}& {\footnotesize 0.078}
%&\ {\footnotesize }
\\
\hline
\end{tabular}

\end{center}
\end{table}

\section{Process Model Cluster Analysis}
In the following section consumable results, actionable information and valuable insights are extracted from all the six clusters obtained using LCS metric. We show that clustering facilitates easier identification of inconsistencies and imperfections and better understanding of the process that would not have been possible by studying the complex spaghetti model.

\begin{table*}
\begin{center}
\footnotesize
\begin{center}
\caption{Self Loops and Back-Forth Analysis }
\label{table:loop}
\end{center}
\begin{tabular}{|c|c|c|c|c|c|c|c|}
\hline
\multicolumn{1}{|c|}{\textbf{{\footnotesize Activity }}} &
\multicolumn{1}{|c|}{\textbf{{\footnotesize Main Model }}} &
\multicolumn{1}{|c|}{\textbf{{\footnotesize Cluster 1 }}} & 
\multicolumn{1}{|c|}{\textbf{{\footnotesize Cluster 2 }}} &
\multicolumn{1}{|c|}{\textbf{{\footnotesize Cluster 3 }}} &
\multicolumn{1}{|c|}{\textbf{{\footnotesize Cluster 4 }}} &
\multicolumn{1}{|c|}{\textbf{{\footnotesize Cluster 5 }}} &
\multicolumn{1}{|c|}{\textbf{{\footnotesize Cluster 6 }}}\\
\hline
{\footnotesize ASS} & {\footnotesize $28$, CCC/15} & {\footnotesize $2$, ATT/1} & {\footnotesize $5$, CCC/1} & {\footnotesize $8$, CCC/6} & {\footnotesize $8$, QAC/3} & {\footnotesize $5$, CCC/7} & {\footnotesize -,-} \\
\hline
{\footnotesize ATT} & {\footnotesize $266$, FLA/116} & {\footnotesize  $24$, FLA/6} & {\footnotesize $20$, FLA/6} & {\footnotesize $102$, FLA/40} & {\footnotesize $48$, FLA/18} & {\footnotesize $70$, FLA/43} & {\footnotesize $2$, FLA/3} \\
\hline
{\footnotesize BLO} & {\footnotesize $152$, CCC/39} & {\footnotesize  $4$, CCC/3} & {\footnotesize $18$, CCC/4} & {\footnotesize $72$, DEP/18} & {\footnotesize $20$, CCC/4} & {\footnotesize $38$, CCC/11} & {\footnotesize -, CCC/2} \\
\hline
{\footnotesize CCC} & {\footnotesize $6776$, FLA/250} & {\footnotesize  $524$, SNR/60}& {\footnotesize $875$, WHI/37} & {\footnotesize $3119$, DEP/141} & {\footnotesize $1345$, SNR/151} & {\footnotesize $871$, FLA/60} & {\footnotesize $42$, SUR/53} \\
\hline
%{\footnotesize CFB} & {\footnotesize $26$, CCC/16} & {\footnotesize $1$, CCC/1}& {\footnotesize $6$, CCC/6} & {\footnotesize $11$, CCC/6} & {\footnotesize -,- } & {\footnotesize $4$, CCC/3} & {\footnotesize $4$,- }  \\
%\hline
%{\footnotesize CFC} & {\footnotesize $1$,-} & {\footnotesize -,-}& {\footnotesize $1$,- } & {\footnotesize -,- } & {\footnotesize -,- } & {\footnotesize -,- } & {\footnotesize -,- }  \\
%\hline

%{\footnotesize CFS }& {\footnotesize $151$, CFT/14} & {\footnotesize $66$, CFT/6}& {\footnotesize $4$,- } & {\footnotesize $14$, SRV/1} & {\footnotesize -,- } & {\footnotesize $66$, CFT/8} & {\footnotesize $1$,- }  \\
%\hline
%{\footnotesize CFT } & {\footnotesize $38$, CCC/15}& {\footnotesize $8$, CFS/3}& {\footnotesize -, CCC/1} & {\footnotesize -, CCC/2} & {\footnotesize $1$,- } & {\footnotesize $29$, CCC/10} & {\footnotesize -,- }  \\
%\hline
{\footnotesize COM} & {\footnotesize $2$, QAC/3} & {\footnotesize -,-} & {\footnotesize -, QAC/1} & {\footnotesize $1$, QAC/1} & {\footnotesize $1$, QAC/1} & {\footnotesize -, CCC/1} & {\footnotesize -,- } \\
\hline
{\footnotesize DEP}& {\footnotesize $704$, CCC/110} & {\footnotesize $9$, SNR/2}& {\footnotesize $21$, CCC/3} & {\footnotesize $576$ , CCC/83} & {\footnotesize $41$, CCC/6} & {\footnotesize $57$, CCC/17} & {\footnotesize $ $, CFL/1} \\
\hline
{\footnotesize FLA}& {\footnotesize $1612$, ATT/464}& {\footnotesize $101$, CCC/32}& {\footnotesize $93$, ATT/25} & {\footnotesize $614$, ATT/168} & {\footnotesize $228$, ATT/48} & {\footnotesize $557$, ATT/186} & {\footnotesize $19$, ATT/12} \\
\hline
%{\footnotesize ISC}& {\footnotesize $1$, SNU/1} & {\footnotesize -, SNU/1}& {\footnotesize -,- } & {\footnotesize $1$, SUN/1} & {\footnotesize -,- } & {\footnotesize -, SUA/1} & {\footnotesize -,- } \\
%\hline
{\footnotesize OPS} & {\footnotesize $1$, PLA/33}& {\footnotesize -,-}& {\footnotesize -, PLA/13 } & {\footnotesize -,PLA/8 } & {\footnotesize $1$,- } & {\footnotesize -, PLA/12 } & {\footnotesize -,- } \\
\hline
%{\footnotesize OPS} & {\footnotesize $1$, PLA/3}& {\footnotesize -,-}& {\footnotesize -, PLA/1 } & {\footnotesize -,- } & {\footnotesize $1$,- } & {\footnotesize -, PLA/2 } & {\footnotesize -,- } \\
%\hline
%    {\footnotesize PRI} & {\footnotesize $3$, TAR/2} & {\footnotesize -, CCC/1}& {\footnotesize -, TAR/2} & {\footnotesize -,- } & {\footnotesize -,- } & {\footnotesize $1$,- } & {\footnotesize $2$,- } \\

%\hline

%{\footnotesize PRO}& {\footnotesize $3$, COM/1} & {\footnotesize -,-}& {\footnotesize $1$,- } & {\footnotesize -, COM/1 } & {\footnotesize $2$,- } & {\footnotesize -,- } & {\footnotesize -,- } \\
%\hline
%{\footnotesize QAC} & {\footnotesize $9$, ASS/7}& {\footnotesize $3$, ASS/1}& {\footnotesize -, COM/1 } & {\footnotesize $1$, ASS/2} & {\footnotesize $5$, ASS/4} & {\footnotesize $1$,- } & {\footnotesize -,-} \\
%\hline
{\footnotesize RES}& {\footnotesize $1$, SRU/2} & {\footnotesize $1$,-}& {\footnotesize -, SRU/1} & {\footnotesize -,- } & {\footnotesize -,- } & {\footnotesize -,- } & {\footnotesize -, SRU/1} \\
\hline
%{\footnotesize SEE} & {\footnotesize $3$,-}& {\footnotesize -,-}& {\footnotesize -,- } & {\footnotesize $1$,- } & {\footnotesize $1$,- } & {\footnotesize $1$,-} & {\footnotesize -,- } \\
%\hline
{\footnotesize SRR} & {\footnotesize -, RFF/6}& {\footnotesize -, RFF/1}& {\footnotesize -,- } & {\footnotesize -, RFF/3} & {\footnotesize -,RES/3 } & {\footnotesize -, RFF/2 } & {\footnotesize -, RES/1 } \\
\hline
{\footnotesize SUM}& {\footnotesize $15$, CCC/27} & {\footnotesize $1$, CCC/2}& {\footnotesize $3$, CCC/1} & {\footnotesize $5$, CCC/11} & {\footnotesize $2$, CCC/12} & {\footnotesize $4$, ASS/1 } & {\footnotesize -,- } \\
\hline
{\footnotesize TAR}& {\footnotesize $21$, CCC/26} & {\footnotesize -, CCC/1}& {\footnotesize $4$, CCC/1} & {\footnotesize $12$, CCC/10} & {\footnotesize $1$, FLA/2} & {\footnotesize $4$, CCC/11} & {\footnotesize -, ASS/1} \\
\hline
{\footnotesize VER}& {\footnotesize $6$, CCC/20} & {\footnotesize $2$,-}& {\footnotesize -, CCC/7 } & {\footnotesize $1$, CCC/7} & {\footnotesize -,PRO/1 } & {\footnotesize $1$,- } & {\footnotesize $2$, CCC/5} \\
\hline
%{\footnotesize WHI} & {\footnotesize $71$, CCC/49}& {\footnotesize $5$, CCC/1}& {\footnotesize $4$,CCC/5 } & {\footnotesize $21$, CCC/16} & {\footnotesize $4$, WHI/5} & {\footnotesize $32$,CCC/12} & {\footnotesize $5$, CCC/10 } \\
%\hline
\end{tabular}
\end{center}
\end{table*}

\subsection{Self-loop Analysis}
Study of self-loops is important since such loops indicate intensive problems \cite{Halverson:2006:DTV:1180875.1180883} which are often difficult to detect because it may seem that at each stage some useful work is being done though actually no progress is being made and the bug is just getting transferred \cite{Halverson:2006:DTV:1180875.1180883}. In a process model, a self-loop can be defined as the transition A$\rightarrow$A i.e. a transition that begins and ends at the same activity. Such anti-patterns are undesirable and cause redundancy in the bug's trace. Just looking at the count of self-loops of an activity in the event log of spaghetti model is not enough since it might happen that most of these self-loops are occurring in traces of a few bugs in which case we cannot generalize and say that this particular activity causes majority of self-loops in the system. Doing self-loop analysis after clustering similar traces helps us to discover if self-loop of an activity appears only in certain kinds of bugs or if it appears in majority in all the traces. First entry in each cell of Table \ref{table:loop} denotes the frequency of self-loop of the activity specified in the first cell of the same row. "-" indicates absence of loop. Due to limited space only some of the activities are represented in Table \ref{table:loop}.

\begin{enumerate}

\item Self-loop frequency of activity Carbon Copy (CCC) is high in all the six clusters with the count being as high as $3119$ in Cluster 3. This indicates that many people including users who have no direct role to play in the bug are added in the mailing list. Its an unhealthy practise to repeatedly add/remove people from the mailing list and should be avoided by adding only a few people who are interested in receiving notifications about the bug's progress.

\item Many self-loops of activity Attachments (ATT: setting attributes of file related to the bug uploaded by a user) in clusters $3$, $4$ \& $5$ indicates that several properties of attachment file\footnote{https://www.bugzilla.org/docs/3.0/html/api/Bugzilla/Attachment.html} associated with a bug like content-type, description, filename, flags etc keep on changing and attribute fields are not correctly entered by the user while filing the bug. 

\item Many recurrent loops of Activity FlagTypes (FLA) occur in Clusters $3$ and $5$. Flags can be of two types: attachment flags and bug flags\footnote{http://www.bugzilla.org/docs/2.22/html/flags-overview.html}. Loop involving the former flag indicates that a developer has asked other developers to review his code implying that peer code review practice is followed for quite a lot of bugs while loop involving the latter type indicates that status information of the bug is repeatedly required e.g needinfo flag is set many times sequentially implying that the developer requires more information about the issue raised indicating that bugs are reported with incomplete information.

\item High Self-Loop frequency of activity Blocks (BLO) in Cluster 3 indicates that several bugs are repeatedly added in the Blocks field which means a lot of bugs are discovered which depends on the current bug. Bugs in this cluster needs to be resolved on a priority basis as several other bugs are dependent on them.

\item Self-loop frequency of activity Depends on (DEP) is extremely high in Cluster 3 indicating that several bugs are identified on which the current bug is dependent. It is interesting to note that self-loop frequency of BLO is also high in this cluster indicating that bugs in these clusters are either dependents or dependees.

\end{enumerate}

\subsection{Back-Forth Analysis}
A back-forth loop, also known as ping pong pattern, can be defined as a transition A$\rightarrow$B$\rightarrow$A i.e. a transition which begins at activity A, goes to activity B and again ends at A. Second entry in each cell of Table \ref{table:loop} contains the activity with which the activity specified in the first cell of the same row is forming a back-forth loop maximum number of times along with the frequency of that loop. An activity A can be in a back-forth loop with multiple activities e.g. A$\rightarrow$B$\rightarrow$A with frequency $f_{1}$ and A$\rightarrow$C$\rightarrow$A with frequency $f_{2}$ and $f_{2}$ $\geq$ $f_{1}$. C/$f_{2}$ is specified as the second entry in the cell corresponding to Activity A in Table \ref{table:loop}. "-" indicates absence of loop. Activities forming loops with high frequency can be effectively analysed in clusters. Since bugs with similar life-cycle are clustered together, root cause behind the occurrence of such anti-patterns also becomes easier to identify and study.
\begin{enumerate}

\item Ping pong patterns that include activity Status Resolved Reopened (SRR) are present in small numbers but are of major interest. The resolve-reopen loop is a problematic pattern. In Clusters $1$, $3$ and $5$ SRR is looping with activity Resolution Fixed (REF) which means that a fixed bug is reopened and again fixed. It happens when the resolution of a resolved bug is found to be incorrect. Such loops are undesirable because the average time to resolve a re-opened bug can be twice as long as the time to resolve a non re-opened bug \cite{emsemad13}. 
\item Activity Depends On (DEP) forms a back-forth loop with Carbon Copy (CCC) $83$ times and CCC forms a loop with DEP $141$ times in Cluster $3$ because the teams solving other bugs on which the current bug is dependent need to be informed about the bug's progress so that they can be included in the decision making process of the current bug. Such loops can be reduced by adding just a few people from other teams in the CC list like the team leader instead of all team members.

\item Important attributes of the bug like version (VER), operating system (OPS), summary (SUM) and target milestone (TAR) are involved in ping pong patterns indicating that it takes repeated efforts to conclude the values of these fields. Bug reporters should be encouraged to write informative summary of the issue and specify fields such as OS and version of the software in which the bug is occurring while filing the bug.

\end{enumerate}

\subsection{Reopen Analysis}
\begin{figure}[t!]
\centering
\noindent\includegraphics[width=9cm,height=7cm]{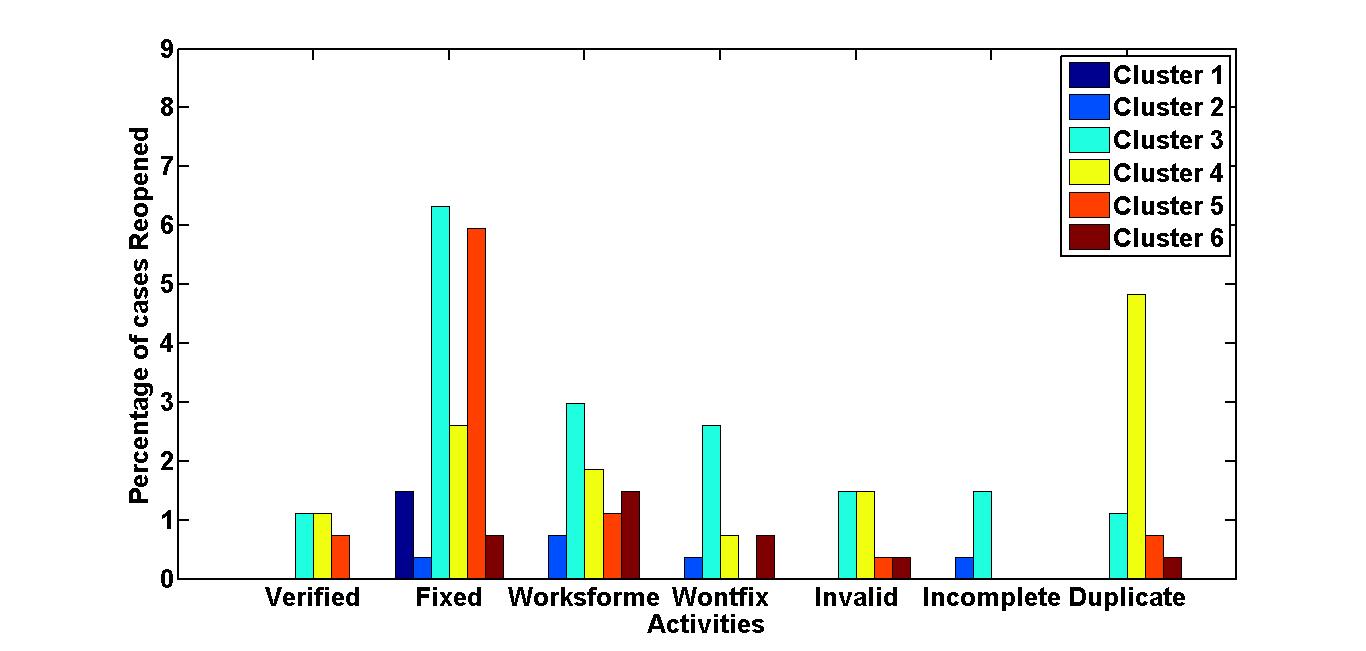}
\caption{Histograms showing percentage of cases Reopened for different states (Activities) across the 6 clusters.}
\label{fig:reopen}
\end{figure}
Bug reopening refers to the act of changing the status of the bug that was once resolved to Reopened (SRR) as the resolution was found to be incorrect forcing the bug to traverse its lifecycle again. Bug reopening is equally important in open source systems like Bugzilla as it is in closed source or commercial systems \cite{export:159352}. It increases the costs of maintaining the software, lessens the user-perceived quality of the system and leads to extra and needless rework by already loaded developers \cite{emsemad13}. Analysis of factors leading to bug reopening helps in improving the quality of bug fixing process and countering all these problems. We take into account the following factors \cite{Gupta:2014:NMB:2590748.2590749} \cite{emsemad13} \cite{export:159352} that contribute in reopening of bugs:
\begin{enumerate}
\item Verified (SRV): A bug verified by a Quality Assurance agent may get reopened if some useful information about the bug becomes available that demands to have it reviewed again.
\item Fixed (REF): A fixed bug may have its reopening if the fix proposed seems to have faults and is not complete and entirely correct solution.
\item Duplicate (RED): If the bug is not studied deeply and few of its symptoms match with some already existing bug, it is incorrectly assumed to be the case of duplicacy.
\item Wontfix (REX)/ Invalid (REN)/ Incomplete (REI)/ Worksforme (REW): There are high chances of re-openings if the developer was not able to fix the bug (Wontfix), issue raised was not categorized as a bug (Invalid), bug was reported with incomplete information (Incomplete) or if it was not successfully reproduced (Worksforme).
\end{enumerate}
We believe that clustering helps in analysing whether the reopening due to an activity is happening globally throughout the main model or in a certain set of similar bugs.
\begin{enumerate}
\item Absence of bug re-opening due to Verified (SRV) in Clusters 1, 2 and 6 is supported by the fact that a Quality Assurance agent (QAC) confirms that a proper fix is achieved. While significant re-opening due to Fixed (REF) in all the clusters especially Clusters $3$ and $5$ indicates bad understanding and management in fixing the bugs in previous releases, leading to loss of time in analysing and correcting the same bug again in the current release (regression bugs). This can be avoided if proper testing and verification of the fix proposed by the developer is done prior to closing the bug.

\item Reopening after activity REW is contributed by $5$ out of $6$ clusters suggesting that re-opening due to Wontfix is occurring globally throughout the dataset and is not limited to some similar types of bugs. Bugs entering into the system are initially difficult to reproduce, thus are left for future references/information using which they will be reopened again. This can be avoided by extracting all possible information about the bug from the reporter to improve understanding before setting its resolution. Also, reporters should be encouraged to describe the bug in as much detail as possible and form for filing a bug should contain various fields that can capture the information about the issue raised in detail.

\item Through clustering we are able to segregate those bugs in Cluster $4$ which get reopened because of incorrectly getting marked as Duplicate (RED) indicating that the bugs are not properly examined before their resolution is set. Process analyst can analyse such bugs to determine whether the duplicacy is due to similar keywords and title used in describing the bug or if the symptoms of the bugs were not studied deeply leading to failure in identification of the root cause of the issue. 

\item One reason behind large number of reopenings due to Worksforme and Wontfix in Cluster $3$ is underestimation of priority of bugs which brings attention to the fact that there is a need to establish clear guidelines and policies to effectively decide priority of a bug.  
\end{enumerate}

\subsection{Bottleneck Identification}

Bottleneck refers to those areas (activities, transitions, paths) of process model that consume comparatively more time than rest of the system causing the entire process to slow down. Identification of principal factors constraining the process speed can help a process analyst in working upon the causes that deter the performance of a process.  
We compute the mean time taken for every transition in main model as well as all the clusters. For analysis, we consider transitions taking maximum amount of time and discover severe bottlenecks present in the models.

\begin{figure}[t]
\centering
\noindent\includegraphics[width=9cm,height=7cm]{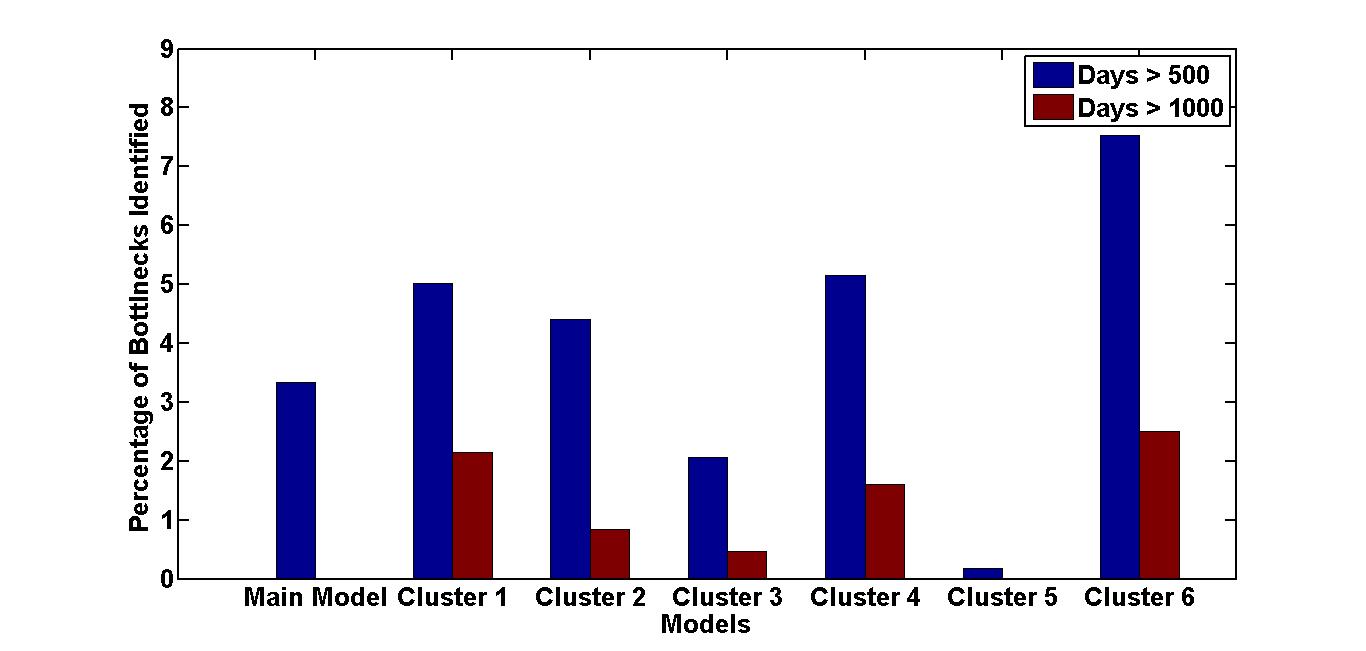}
\caption{Histograms showing percentage of Bottlenecks identified in the Main Model and 6 clusters.}
\label{fig:bottleneck}
\end{figure}

\begin{enumerate}
\item Figure \ref{fig:bottleneck} shows percentage of bottlenecks taking mean time more than $500$ and $1000$ days in main model and the 6 clusters. From Figure \ref{fig:bottleneck}, we observe that percentage of bottlenecks taking more than 500 days (mean value) is greater in Clusters $1$, $2$, $4$ and $6$ as compared to the main model. While for duration greater than $1000$ days (mean value), each cluster has higher percentage of bottlenecks than the main model. It is due to the absolute count of transitions which is less in a cluster than the main model producing greater mean value for the clusters. Thus, bottlenecks that are not quite evident in the main model are clearly visible in the clusters.
\item Set of transitions, taking mean time greater than 1000 days, found in both the main model as well as clusters are:
\begin{enumerate}
\item SRV $\to$ CFB, SRV $\to$ QAC  implying that after a bug is verified (SRV), there is a large gap before any other actions like contacting Quality Assurance agent (QAC) and setting any blocking flag (CFB) are done. This indicates that once a bug is verified it is not acted upon much.

\item ISC $\to$ SNR suggesting huge delay between the time a bug is confirmed to be true (ISC) to the time appropriate actions are taken to resolve it (SNR) indicating that in some cases it takes a lot of time to understand and confirm that the issue raised is actually a bug. This confirmation step (ISC) can be accelerated to make the system faster.

\end{enumerate}
\item The bottlenecks found in clusters (not observed in main model) taking mean time greater than $1000$ days are:
\begin{enumerate}
\item Some of the activities performed before changing the status of bug from New to Resolved (SNR) like ASS and ATT take over $1200$ days suggesting large delays in assigning the bug to a resolver (ASS) and studying the associated attachments (ATT). Many a times the attachments are obsolete and their attributes are not defined properly leading to wastage of time in asking the user to upload the attachment again and resetting their attribute values. Also, delay in assigning the bug to a developer needs to be removed by having a proper procedure to quickly select an appropriate resolver for the bug. 

\item Our analysis shows that setting the resolution to Incomplete, Worksforme and Wontfix takes a lot of time as transitions CCC $\to$ REI, CCC $\to$ REW, CCC $\to$ REX are taking more than $3$ years. Higher efficiency is required to identify such cases so that their resolution can be set quickly. Reopening of bugs with these resolutions also takes considerable amount of time indicating that reasons of reopening due to these factors need to be studied in detail with higher priority.

\item Changing the status of bug from Unconfirmed to Resolved is taking $4$ years (indicated by transitions SUM $\to$ SUR, OPS $\to$ SUR) because important attributes of bug like summary (SUM) and operating system (OPS) were not properly defined by the bug reporter, so determining their values took a lot of time. 
\end{enumerate}
\end{enumerate}

\section{Determining the Best Cluster Solution}
Clustering can give many different solutions depending upon the algorithm used, initial cluster centers chosen, number of iterations and number of clusters specified. Out of the many possible solutions, we select the one where clusters have low complexity and high fitness value for enabling better analysis.  To test the proposed algorithm, experimental dataset described in Table \ref{table:dataset_details} is split into four equal sub datasets and each subset is experimented with the proposed algorithm using $k$-medoid with LCS similarity metric. Algorithm \ref{automate} runs the clustering algorithm thrice over the input event log to select the best cluster set. Table \ref{table:ac} gives the G\_Ratio of all the three iterations performed on all four sub datasets as well as the iteration whose solution set is determined to be the best by our proposed algorithm.

\begin{algorithm}
%\dontprintsemicolon
\KwData{\footnotesize{History data of bugs}}
\KwResult{\footnotesize{Best cluster set}}
\textnormal{\footnotesize{Perform the preprocessing steps and obtain the sequential data format from the history of bugs extracted}}\\
\textnormal{\footnotesize{generate $3$ cluster sets $S_{1}$, $S_{2}$ and $S_{3}$ using $k$-Medoid Clustering that uses LCS/DTW similarity for input $k$ value}}\\
\ForEach {\textnormal{\footnotesize{cluster set $S_{i}$ consisting of $m$ clusters}}}{

\For{\textnormal{\footnotesize j}$\leftarrow1 $ \KwTo \footnotesize m }{
\footnotesize {discover process model $P_{j}$}
}
\vspace{-3pt}
{\footnotesize
\begin{flalign*}
  {C\_score_{i}} &= \sum\limits_{{j=1}}^m \frac{\parbox{1.35in}{$ {Complexity}(  {Xml}\ format\\$\hspace{0.cm}$ input\ of\ P_{j})*t_j$}}{t_j}\\
   F\_score_{i} &= \sum\limits_{j=1}^m \frac{\parbox{1.65in}{$Fitness(Xml\ format\ input\ of\\
 $\hspace{0.cm}$P_{j},\ Eventlog\ in\ sequential\\$\hspace{0.cm}$ format\ of\ cluster\ C_{j})*t_j$}}{t_j}\\
 G\_Ratio_{i} &= F\_score/C\_score
 \end{flalign*}
 }%
 \textnormal{\footnotesize{where $t_j$ is total traces in event log of cluster $C_j$}}
}
\textnormal{\footnotesize{return the cluster set $S_{i}$ with the maximum G\_Ratio}}.\\
\caption{Automate Clustering }
\label{automate}
\end{algorithm}

\begin{table}[!htp]
\begin{center}
\normalsize
\begin{center}
\caption{Automated Clustering Algorithm Analysis }
\label{table:ac}
\end{center}
\begin{tabular}{|c|c|c|c|c|c|}
\hline
\multicolumn{1}{|>{\centering\arraybackslash}m{0.8 cm}|}{\centering \textbf{\footnotesize Dataset} } &
\multicolumn{1}{|m{0.75 cm}|}{\centering \textbf{ \footnotesize Iteration} } &
\multicolumn{1}{|m{1 cm}|}{\centering \textbf{\footnotesize Weighted \footnotesize Complexity} } &
\multicolumn{1}{|m{1 cm}|}{\centering \textbf{\footnotesize Weighted \footnotesize Fitness} } &
\multicolumn{1}{|m{2 cm}|}{\centering \textbf{\footnotesize{ G\_Ratio}} }&
\multicolumn{1}{|>{\centering\arraybackslash}m{0.8 cm}|}{\centering \textbf{\footnotesize Result}}\\
\hline
$1$ & $1$  & {$90.98$} & {$0.190$} & {$2.08 \times 10^{-03}$} &{-}\\
\hline
$1$ & $2$  & {$92.38$} & {$0.158$} & {$1.71 \times 10^{-03}$} &{-}\\
\hline
$1$ & $3$  & {$90.91$} & {$0.227$} & {$2.49 \times 10^{-03}$}&{\footnotesize{Selected}}\\
\hline
$2$ & $1$ & {$99.43$} & {$0.275$} & {$2.08 \times 10^{-03}$}&{-}\\
\hline
$2$ & $2$ & {$100.99$} & {$0.205$} & {$2.7 \times 10^{-03}$}&{\footnotesize{Selected}}\\
\hline
$2$ & $3$ & {$105.8$} & {$0.213$} & {$2.01 \times 10^{-03}$} &{-}\\
\hline
$3$ & $1$ & {$92.05$} & {$0.125$} & {$1.35 \times 10^{-03}$}&{-}\\
\hline
$3$ & $2$ & {$91.39$} & {$0.106$} & {$1.15 \times 10^{-03}$}&{-}\\
\hline
$3$ & $3$ & {$93.47$} & {$0.218$} & {$2.33 \times 10^{-03}$}&{\footnotesize{Selected}}\\
\hline
$4$ & $1$ & {$81.36$} & {$0.394$} & {$4.84 \times 10^{-03}$}&{\footnotesize{Selected}}\\
\hline
$4$ & $2$ & {$85.57$} & {$0.270$} & {$3.15 \times 10^{-03}$}&{-}\\
\hline
$4$ & $3$ & {$85.40$} & {$0.211$} & {$2.47 \times 10^{-03}$}&{-}\\
\hline
\end{tabular}
\end{center}
\end{table}

\section{Related Work and Research Contributions}
Real life event logs are diverse, unstructured and complex leading to formation of 'Spaghetti Models'. The problem of spaghetti process models has been discussed in \cite{DBLP:conf/acsd/AalstG07} and \cite{veiga10understanding}. Several techniques have been proposed in literature to cluster traces to deal with complex process models. Bose et al. propose a context aware approach to cluster process instances based on Levenshtein distance \cite{DBLP:conf/sdm/BoseA09}. In the technique substitution, insertion and deletion costs of symbols are derived for similarity. The authors evaluate the proposed algorithm on the telephone repair process event log and show that the approach is able to generate clusters with high degree of fitness and comprehensibility when compared to other approaches \cite{DBLP:conf/sdm/BoseA09}. In
\cite{DBLP:conf/acsd/AalstG07} Aalst et al. apply combination of abstraction and clustering techniques to simplify spaghetti-like models discovered using process mining techniques from unstructured and complicated processes \cite{DBLP:conf/acsd/AalstG07}. They use significance and correlation metrics to simplify the processes by clustering less significant but highly correlated data \cite{DBLP:conf/acsd/AalstG07}. Ferreira et al. propose a sequence clustering approach where each cluster is represented by a first-order Markov chain. \cite{Ferreira:2007:APM:1793114.1793147}. Veiga et al. extended this work by   using two dummy states (input and output state) with the Markov chain model for depicting the probability for an event to be the first or last in the sequence \cite{veiga10understanding}. They also suggest several preprocessing steps done before clustering to eliminate undesirable events from the event log \cite{veiga10understanding}. Weerdt et al. propose a new tecnique called ActiTraC (active trace clustering) for trace clustering which uses elements of active learning in an unsupervised environment \cite{DBLP:journals/tkde/WeerdtBVB13}. The proposed algorithm lessens the divergence between the clustering bias and the evaluation bias and improves the accuracy and complexity of process models \cite{DBLP:journals/tkde/WeerdtBVB13}. Song et al. \cite{DBLP:conf/bpm/SongGA08} propose a trace clustering technique that uses several perspectives of traces such as performance, transition, case and event attributes organised as a feature vector. Conformance measurement done through process mining in business processes has been shown in \cite{annebpmworkshop}, \cite{DeltaConform} and \cite{Gupta:2014:NMB:2590748.2590749}.
In context to existing work, the  paper makes the following novel contributions:
\begin{enumerate}
\item Improving the goodness (complexity and fitness) of process models by splitting the event-log into homogeneous subsets by clustering structurally similar traces by adapting the the $K$-Medoid algorithm.
\item Use of Longest Common Subsequence (LCS) and Dynamic Time Warping (DTW) distance metrics in the adaptation of $K$-medoid algorithm.
\item Illustrating the benefits of trace clustering in identifying bottlenecks and study of back-forth \& self-loops and bug reopening.
\item An algorithm to automate clustering that returns the best cluster set for an event log by determining the goodness of process models.
\item An in-depth case study on the open source Firefox browser project to investigate the effectiveness of the proposed approach.
\end{enumerate}

\section{Conclusion}
Analysing the results after mining real world unstructured event logs that show adhoc behavior is difficult due to production of complex spaghetti-like process models. Our work is a contribution towards simplifying these complex models by means of clustering so that they can be easily understood by the process analyst. We adapt $K$-medoid algorithm using two different distance metrics- LCS and DTW to obtain clusters having good intra-class similarity. $K$-medoid is an efficient clustering algorithm which is insensitive to outliers and noisy data. Goodness of the models increase as fitness and structural complexity is improved making the models easier to comprehend. We demonstrate the effectiveness of our proposed technique by performing a real life case study on Firefox browser project. We successfully show that clustering enables better analysis, making it easier to identify bottlenecks, study reopening of bugs, self \& back forth loops. We propose an algorithm that returns the cluster set with highest goodness ratio to automate the clustering process which is effectively tested on four different datasets.
%\nocite{*}
\bibliographystyle{IEEEtran}
\bibliography{citation}

\begin{thebibliography}{10}
\providecommand{\url}[1]{#1}
\csname url@samestyle\endcsname
\providecommand{\newblock}{\relax}
\providecommand{\bibinfo}[2]{#2}
\providecommand{\BIBentrySTDinterwordspacing}{\spaceskip=0pt\relax}
\providecommand{\BIBentryALTinterwordstretchfactor}{4}
\providecommand{\BIBentryALTinterwordspacing}{\spaceskip=\fontdimen2\font plus
\BIBentryALTinterwordstretchfactor\fontdimen3\font minus
  \fontdimen4\font\relax}
\providecommand{\BIBforeignlanguage}[2]{{%
\expandafter\ifx\csname l@#1\endcsname\relax
\typeout{** WARNING: IEEEtran.bst: No hyphenation pattern has been}%
\typeout{** loaded for the language `#1'. Using the pattern for}%
\typeout{** the default language instead.}%
\else
\language=\csname l@#1\endcsname
\fi
#2}}
\providecommand{\BIBdecl}{\relax}
\BIBdecl

\bibitem{poncin2011process}
W.~Poncin, A.~Serebrenik, and M.~van~den Brand, ``Process mining software
  repositories,'' in \emph{Software Maintenance and Reengineering (CSMR), 2011
  15th European Conference on}.\hskip 1em plus 0.5em minus 0.4em\relax IEEE,
  2011, pp. 5--14.

\bibitem{rubin2007process}
V.~Rubin, C.~W. G{\"u}nther, W.~M. Van Der~Aalst, E.~Kindler, B.~F. Van~Dongen,
  and W.~Sch{\"a}fer, ``Process mining framework for software processes,'' in
  \emph{Software Process Dynamics and Agility}.\hskip 1em plus 0.5em minus
  0.4em\relax Springer, 2007, pp. 169--181.

\bibitem{Gunther:2007:FMA:1793114.1793145}
\BIBentryALTinterwordspacing
C.~W. G\"{u}nther and W.~M.~P. Van Der~Aalst, ``Fuzzy mining: Adaptive process
  simplification based on multi-perspective metrics,'' in \emph{Proceedings of
  the 5th International Conference on Business Process Management}, ser.
  BPM'07.\hskip 1em plus 0.5em minus 0.4em\relax Berlin, Heidelberg:
  Springer-Verlag, 2007, pp. 328--343. [Online]. Available:
  \url{http://dl.acm.org/citation.cfm?id=1793114.1793145}
\BIBentrySTDinterwordspacing

\bibitem{kaufman1987clustering}
\BIBentryALTinterwordspacing
L.~Kaufman and P.~Rousseeuw, \emph{Clustering by Means of Medoids}, ser.
  Reports of the Faculty of Mathematics and Informatics.\hskip 1em plus 0.5em
  minus 0.4em\relax Faculty of Mathematics and Informatics, 1987. [Online].
  Available: \url{http://books.google.co.in/books?id=HK-4GwAACAAJ}
\BIBentrySTDinterwordspacing

\bibitem{kmedbook90}
\BIBentryALTinterwordspacing
L.~Kaufman and P.~J. Rousseeuw, \emph{{Finding groups in data: an introduction
  to cluster analysis}}.\hskip 1em plus 0.5em minus 0.4em\relax John Wiley \&
  Sons, Inc., 1990. [Online]. Available:
  \url{http://books.google.com/books?id=yS0nAQAAIAAJ}
\BIBentrySTDinterwordspacing

\bibitem{Wagner:1974:SCP:321796.321811}
\BIBentryALTinterwordspacing
R.~A. Wagner and M.~J. Fischer, ``The string-to-string correction problem,''
  \emph{J. ACM}, vol.~21, no.~1, pp. 168--173, Jan. 1974. [Online]. Available:
  \url{http://doi.acm.org/10.1145/321796.321811}
\BIBentrySTDinterwordspacing

\bibitem{Bergroth:2000:SLC:829519.830817}
\BIBentryALTinterwordspacing
L.~Bergroth, H.~Hakonen, and T.~Raita, ``A survey of longest common subsequence
  algorithms,'' in \emph{Proceedings of the Seventh International Symposium on
  String Processing Information Retrieval (SPIRE'00)}, ser. SPIRE '00.\hskip
  1em plus 0.5em minus 0.4em\relax Washington, DC, USA: IEEE Computer Society,
  2000, pp. 39--. [Online]. Available:
  \url{http://dl.acm.org/citation.cfm?id=829519.830817}
\BIBentrySTDinterwordspacing

\bibitem{Hirschberg:1977:ALC:322033.322044}
\BIBentryALTinterwordspacing
D.~S. Hirschberg, ``Algorithms for the longest common subsequence problem,''
  \emph{J. ACM}, vol.~24, no.~4, pp. 664--675, Oct. 1977. [Online]. Available:
  \url{http://doi.acm.org/10.1145/322033.322044}
\BIBentrySTDinterwordspacing

\bibitem{1275947}
J.~Kruskal and M.~Liberman, ``{The symmetric time-warping problem: from
  continuous to discrete},'' 1983.

\bibitem{conf/kdd/BerndtC94}
\BIBentryALTinterwordspacing
D.~J. Berndt and J.~Clifford, ``Using dynamic time warping to find patterns in
  time series.'' in \emph{KDD Workshop}, U.~M. Fayyad and R.~Uthurusamy,
  Eds.\hskip 1em plus 0.5em minus 0.4em\relax AAAI Press, 1994, pp. 359--370.
  [Online]. Available:
  \url{http://dblp.uni-trier.de/db/conf/kdd/kdd94.html#BerndtC94}
\BIBentrySTDinterwordspacing

\bibitem{books/daglib/0018897}
G.~Gan, C.~Ma, and J.~Wu, \emph{Data clustering - theory, algorithms, and
  applications.}\hskip 1em plus 0.5em minus 0.4em\relax SIAM, 2007.

\bibitem{Cardoso:2006:DCP:2135571.2135586}
\BIBentryALTinterwordspacing
J.~Cardoso, J.~Mendling, G.~Neumann, and H.~A. Reijers, ``A discourse on
  complexity of process models,'' in \emph{Proceedings of the 2006
  International Conference on Business Process Management Workshops}, ser.
  BPM'06.\hskip 1em plus 0.5em minus 0.4em\relax Berlin, Heidelberg:
  Springer-Verlag, 2006, pp. 117--128. [Online]. Available:
  \url{http://dx.doi.org/10.1007/11837862_13}
\BIBentrySTDinterwordspacing

\bibitem{McCabe:1976:CM:800253.807712}
\BIBentryALTinterwordspacing
T.~J. McCabe, ``A complexity measure,'' in \emph{Proceedings of the 2Nd
  International Conference on Software Engineering}, ser. ICSE '76.\hskip 1em
  plus 0.5em minus 0.4em\relax Los Alamitos, CA, USA: IEEE Computer Society
  Press, 1976, pp. 407--. [Online]. Available:
  \url{http://dl.acm.org/citation.cfm?id=800253.807712}
\BIBentrySTDinterwordspacing

\bibitem{annebpmworkshop}
A.~Rozinat and W.~Aalst, ``{Conformance Testing: Measuring the Fit and
  Appropriateness of Event Logs and Process Models},'' in \emph{{First
  International Workshop on Business Process Intelligence (BPI'05)}},
  M.~Castellanos and T.~Weijters, Eds., Nancy, France, September 2005, pp.
  1--12.

\bibitem{Gupta:2014:NMB:2590748.2590749}
\BIBentryALTinterwordspacing
M.~Gupta and A.~Sureka, ``Nirikshan: Mining bug report history for discovering
  process maps, inefficiencies and inconsistencies,'' in \emph{Proceedings of
  the 7th India Software Engineering Conference}, ser. ISEC '14.\hskip 1em plus
  0.5em minus 0.4em\relax New York, NY, USA: ACM, 2014, pp. 1:1--1:10.
  [Online]. Available: \url{http://doi.acm.org/10.1145/2590748.2590749}
\BIBentrySTDinterwordspacing

\bibitem{Halverson:2006:DTV:1180875.1180883}
\BIBentryALTinterwordspacing
C.~A. Halverson, J.~B. Ellis, C.~Danis, and W.~A. Kellogg, ``Designing task
  visualizations to support the coordination of work in software development,''
  in \emph{Proceedings of the 2006 20th Anniversary Conference on Computer
  Supported Cooperative Work}, ser. CSCW '06.\hskip 1em plus 0.5em minus
  0.4em\relax New York, NY, USA: ACM, 2006, pp. 39--48. [Online]. Available:
  \url{http://doi.acm.org/10.1145/1180875.1180883}
\BIBentrySTDinterwordspacing

\bibitem{emsemad13}
E.~Shihab, A.~Ihara, Y.~Kamei, W.~M. Ibrahim, M.~Ohira, B.~Adams, A.~E. Hassan,
  and K.~ichi Matsumoto, ``Studying re-opened bugs in open source software,''
  G.~Antoniol and M.~Pinzger, Eds.\hskip 1em plus 0.5em minus 0.4em\relax
  Springer, 2013, vol.~18, pp. 1005--1042.

\bibitem{export:159352}
\BIBentryALTinterwordspacing
T.~Zimmermann, N.~Nagappan, P.~J. Guo, and B.~Murphy, ``Characterizing and
  predicting which bugs get reopened,'' in \emph{Proceedings of the 34th
  International Conference on Software Engineering (ICSE 2012 SEIP
  Track)}.\hskip 1em plus 0.5em minus 0.4em\relax IEEE, June 2012. [Online].
  Available:
  \url{http://research.microsoft.com/apps/pubs/default.aspx?id=159352}
\BIBentrySTDinterwordspacing

\bibitem{DBLP:conf/acsd/AalstG07}
\BIBentryALTinterwordspacing
W.~M.~P. van~der Aalst and C.~W. G{\"{u}}nther, ``Finding structure in
  unstructured processes: The case for process mining,'' in \emph{Seventh
  International Conference on Application of Concurrency to System Design
  {(ACSD} 2007), 10-13 July 2007, Bratislava, Slovak Republic}, 2007, pp.
  3--12. [Online]. Available:
  \url{http://doi.ieeecomputersociety.org/10.1109/ACSD.2007.50}
\BIBentrySTDinterwordspacing

\bibitem{veiga10understanding}
G.~M. Veiga and D.~R. Ferreira, ``Understanding spaghetti models with sequence
  clustering for {ProM},'' in \emph{Business Process Management Workshops, BPM
  2009 International Workshops, Ulm, Germany, September 7, 2009. Revised
  Papers}, ser. Lecture Notes in Business Information Processing,
  vol.~43.\hskip 1em plus 0.5em minus 0.4em\relax Springer, 2010, pp. 92--103.

\bibitem{DBLP:conf/sdm/BoseA09}
\BIBentryALTinterwordspacing
R.~P. J.~C. Bose and W.~M.~P. van~der Aalst, ``Context aware trace clustering:
  Towards improving process mining results,'' in \emph{Proceedings of the
  {SIAM} International Conference on Data Mining, {SDM} 2009, April 30 - May 2,
  2009, Sparks, Nevada, {USA}}, 2009, pp. 401--412. [Online]. Available:
  \url{http://dx.doi.org/10.1137/1.9781611972795.35}
\BIBentrySTDinterwordspacing

\bibitem{Ferreira:2007:APM:1793114.1793147}
\BIBentryALTinterwordspacing
D.~Ferreira, M.~Zacarias, M.~Malheiros, and P.~Ferreira, ``Approaching process
  mining with sequence clustering: Experiments and findings,'' in
  \emph{Proceedings of the 5th International Conference on Business Process
  Management}, ser. BPM'07.\hskip 1em plus 0.5em minus 0.4em\relax Berlin,
  Heidelberg: Springer-Verlag, 2007, pp. 360--374. [Online]. Available:
  \url{http://dl.acm.org/citation.cfm?id=1793114.1793147}
\BIBentrySTDinterwordspacing

\bibitem{DBLP:journals/tkde/WeerdtBVB13}
\BIBentryALTinterwordspacing
J.~D. Weerdt, S.~K. L.~M. vanden Broucke, J.~Vanthienen, and B.~Baesens,
  ``Active trace clustering for improved process discovery,'' \emph{{IEEE}
  Trans. Knowl. Data Eng.}, vol.~25, no.~12, pp. 2708--2720, 2013. [Online].
  Available: \url{http://doi.ieeecomputersociety.org/10.1109/TKDE.2013.64}
\BIBentrySTDinterwordspacing

\bibitem{DBLP:conf/bpm/SongGA08}
\BIBentryALTinterwordspacing
M.~Song, C.~W. G{\"{u}}nther, and W.~M.~P. van~der Aalst, ``Trace clustering in
  process mining,'' in \emph{Business Process Management Workshops, {BPM} 2008
  International Workshops, Milano, Italy, September 1-4, 2008. Revised Papers},
  2008, pp. 109--120. [Online]. Available:
  \url{http://dx.doi.org/10.1007/978-3-642-00328-8_11}
\BIBentrySTDinterwordspacing

\bibitem{DeltaConform}
\BIBentryALTinterwordspacing
W.~Aalst, ``\BIBforeignlanguage{English}{Business alignment: using process
  mining as a tool for delta analysis and conformance testing},''
  \emph{\BIBforeignlanguage{English}{Requirements Engineering}}, vol.~10,
  no.~3, pp. 198--211, 2005. [Online]. Available:
  \url{http://dx.doi.org/10.1007/s00766-005-0001-x}
\BIBentrySTDinterwordspacing

\end{thebibliography}

\end{document}